# Constrained Generative Adversarial Network Ensembles for Sharable Synthetic Data Generation

Engin Dikici, Luciano M. Prevedello, Matthew Bigelow, Richard D. White, and Barbaros Selnur Erdal

## Abstract

The sharing of medical imaging datasets between institutions, and even inside the same institution, is limited by various regulations/legal barriers. Although these limitations are necessities for protecting patient privacy and setting strict boundaries for data ownership, medical research projects that require large datasets suffer considerably as a result. Machine learning has been revolutionized with the emerging deep neural network approaches over recent years, making the data-related limitations even a larger problem as these novel techniques commonly require immense imaging datasets. This paper introduces constrained Generative Adversarial Network ensembles (cGANe) to address this problem by altering the representation of the imaging data, whereas containing the significant information, enabling the reproduction of similar research results elsewhere with the sharable data. Accordingly, a framework representing the generation of a cGANe is described, and the approach is validated for the generation of synthetic 3D brain metastatic region data from T1-weighted contrast-enhanced MRI studies. For 90% brain metastases (BM) detection sensitivity, our previously reported detection algorithm produced on average 9.12 false-positive BM detections per patient after training with the original data, whereas producing 9.53 false-positives after training with the cGANe generated synthetic data. Although the applicability of the introduced approach needs further validation studies with a range of medical imaging data types, the results suggest that the BM-detection algorithm can achieve comparable performance by using cGANe generated synthetic data. Hence, the generalization of the proposed approach for various modalities may occur in the near future.

## 1. Introduction

The amount of training data may be a limiting factor in determining the performance of a machine learning (ML)-based system; especially if the used ML algorithm is highly parametric such as a deep neural network (DNN) [1]. Furthermore, the variability of the training data has a high impact on the generalizability of the solution [2]. Medical imaging systems, deploying DNNs among other ML methodologies [3,4], are highly exposed to these data-related issues [5].

At the same time, patient privacy, data ownership, and regulations (e.g. HIPPAA) limit the usage/sharing of medical data between the institutions, and even between the departments within the same institution [6]. While there are multiple successful initiatives for aggregating multi-institutional public datasets [7–

9], access to large-scale datasets collected from specific modalities for specific medical conditions is not always possible [10].

Nevertheless, restrictions on data-sharing for various application domains have been overcome by the use of collaborated learning (CL), including various methodologies such as federated learning (FL) [11], split learning [12,13], cyclical weight updates (CWU) [14], and large-scale synchronous gradient descent [15]. Briefly, the CL-based approaches facilitate the training of versions of a common ML model at various sites, without the requirement for sites to share data. The application of CL in medical image analysis has been relatively recent. In [16], the brain tumor segmentation model for magnetic resonance imaging (MRI) was deployed and maintained in a multi-institutional format using FL; the study reported that the accuracy of the model, computed via the Dice metric, was comparable with a model trained based on all data. Li et al. [17] investigated a similar problem; however, they focused on possible variations of FL and their impacts on the model performance. In [12], Poirot et al. showed the application of split learning in both the multi-label classification of chest X-ray images and binary classification of diabetic retinopathy images. The authors concluded that the number of sites/institutions do not cause any negative impact on the classification accuracy for these medical imaging scenarios. In their work [14], Chang et al. simulated the dissemination of DNNs for the binary classification of diabetic retinopathy images across four institutions, where the model weights updated via CWU; the study showed that the performance of the distributed model performance was almost identical to that if the imaging data was centrally hosted.

Generative adversarial networks (GANs) [18] have opened up new horizons for the processing of any-dimensional data in various domains (e.g. computer vision, natural language processing, time-series processing, etc.) by unveiling the immense potential of using adversarial loss during the optimization process [19]. Accordingly, GANs have been utilized in medical imaging for various conventional tasks (i.e. image reconstruction, segmentation, classification, detection, and registration), and more recently for medical image synthesis [20]. The medical image synthesis supported by GANs takes advantage of the generative aspect of this methodology; it allows the generation of novel image datasets representing the probability distribution of the original data [21], ideally without replicating or closely mimicking it. This feature was previously exploited to increase the sizes of smaller medical image datasets via augmentation of original data with GAN-generated synthetic data [22–26]. However, as reported previously [23], the inclusion of a high percentage of the original data was needed to achieve acceptable performance in these studies; showing that the augmented data via a GAN could not solely replace the original data.

The motivation for our study is to provide a novel data-sharing protocol that addresses the shortcomings of the CL approaches via a framework introducing the ensemble of constrained GANs (cGANe). We argue that the cGANe, in which: (1) the components (i.e., base-learners) are constrained for their generative abilities, and (2) the size of an ensemble is controlled based on a validation model, has potential to generate viable synthetic data. This allows: (1) a comparable validation model using research results to be reproduced by the client sites, and/or (2) the receiving sites to perform their novel research with shared synthetic data. In practice, the owner of the original data generates a cGANe for its data, altering the representation of information in the datasets to allow comparable research, and then shares the representative synthetic data with other sites. Therefore, unlike the CL approaches, collaborating sites are not required to maintain a common/agreed ML model, and ML models are not shared partially or fully between sites. The overall structure of our proposed framework is illustrated in Figure 1.

This report specifically represents the applicability of the cGANe approach to the generation of 3D brain metastatic-region data from T1-weighted contrast-enhanced MRI studies. First, the components of the framework are introduced. Secondly, the used 3D deep convolutional GAN (DCGAN) formulation is described, and the usage of an ensemble of GANs for addressing the shortcomings of a single GAN is justified. As the performance of an ensemble relies on its components, the Fréchet Distance (FD)-based GAN ensemble growth strategy is then introduced. Our previously described brain metastases (BM)-detection framework [27] is used for validating the generative aspect of a given ensemble in this study; hence, this framework is briefly described. Finally, a validation study that compares the performances of the BM-detection framework trained with the original and cGANe generated synthetic data is provided. The report concludes with: (1) a discussion of the results, (2) a summary of the impact of this study, and (3) a description of limitations and directions for future work.

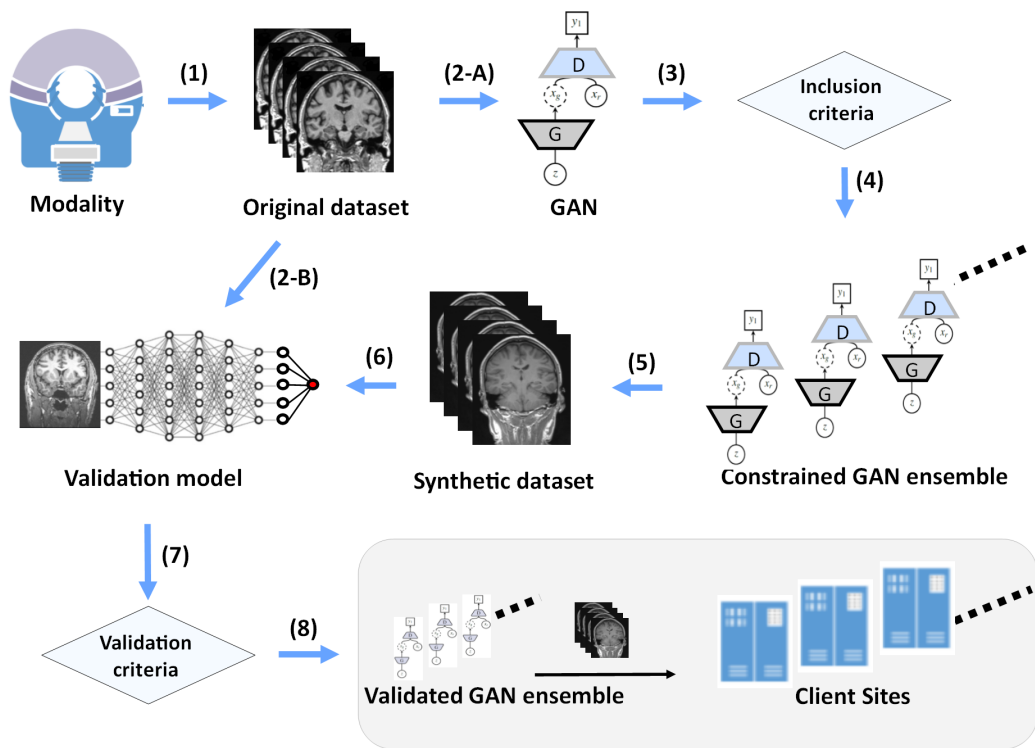

Fig. 1 Constrained GAN ensemble for data sharing.

## 2. Materials and Methods

This study introduces the cGANe concept to generate synthetic medical imaging data that can be shared with the other institutions/sites. The applicability of the proposed approach is demonstrated for a case study where the institution: (1) owns a large set of 3D brain metastatic regions segmented from gadolinium-enhanced T1-weighted MRI datasets, and (2) intends to share this imaging data with other institutions that may benefit from it. However, direct sharing of this specialized dataset may not be an option due to the aforementioned restrictions; thus, the cGANe is set as:

- Medical imaging data reflects segmented 3D brain metastatic foci from a T1-weighted contrast-enhanced MRI; the study demonstrates procedural steps for generating a synthetic version of this data type.
- DCGAN formulation from [28] is adapted for generating the synthetic data.
- The FD [29], computed between the sample set of synthetic data generated by DCGAN and the original data, is utilized as the inclusion criterion. Based on the FD, a given DCGAN can be added to the cGANe.
- A synthetic dataset generated by a cGANe is validated using a BM-detection framework [27], which is the validation model for this case study. The validation criterion is to achieve a comparable detection performance for a preset sensitivity percentage by using the positive synthetic data samples. The threshold, referred to as "baseline-performance" throughout the manuscript, is determined by executing the validation model with the original data. Satisfying this criterion denotes that the cGANe population is adequate. Otherwise, the cGANe should continue to grow with additional GANs.
- A validated cGANe is capable of producing synthetic volumetric BM datasets that can be shared with other sites.

The following subsections provide further algorithmic details for these components.

2.1 Volumetric BM Region Generation with GANs

The generic GAN [18] is a generative model consisting of two neural networks: (1) the generator network producing the synthetic data, and (2) the discriminator network classifying a given data as synthetic or real. Iterative and simultaneous training of these networks is possible by deploying a two-player minimax game; one of the objective functions maximizes the discrimination accuracy, and the other minimizes the synthetic data's correct classification probability:

$$min_G max_D V(D, G) = E_{x \sim p_{data}}[\log D(x)] + E_{z \sim p_{noise}}\left[\log\left(1 - D(G(z))\right)\right], \qquad (1)$$

where (1) $D$ and $G$ are the discriminator and synthetic data generation models, (2) $p_{data}$ is the unknown probability distribution function (PDF) for the real data, and (3) $p_{noise}$ is the PDF for the generator's noise type input (typically uniform or Gaussian).

The network architecture and the optimization criteria (set via loss function) determine the quality and diversity of the synthetic data generated by a GAN; thus, there are various GAN-variants [30]. For the unconditional synthesis of medical images (i.e., generation of medical images from noise vector without any other conditional information) DCGAN, Wasserstein GAN (WGAN) [31] and Progressive Growing GAN (PGGAN) [32] have been the most commonly used formulations due to their good training stabilities [20].

In this study, 3D brain metastatic regions segmented from T1-weighted contrast-enhanced MRI are synthesized by the modified version of DCGAN. The formulation is chosen due to prior studies showing its applicability in the given domain [22,25,33,34], and its ease of implementation in 3D. The original DCGAN, proposed by Radford et al. [28], is adapted for our case by (1) modifying the generator to produce $16 \times 16 \times 16$ volumes (instead of $64 \times 64$ images) that represent cropped BM regions, and (2) modifying the discriminator to classify these volumetric inputs. The implemented DCGAN architecture is given in Figure 2, and some examples for the real and DCGAN generated synthetic BM samples are shown in Figure 3.

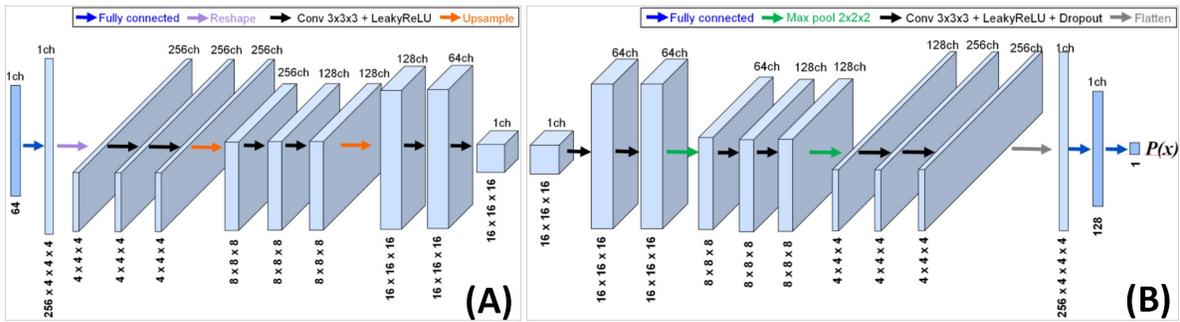

Fig. 2 The generator (A) and discriminator (B) networks of the used 3D DCGAN.

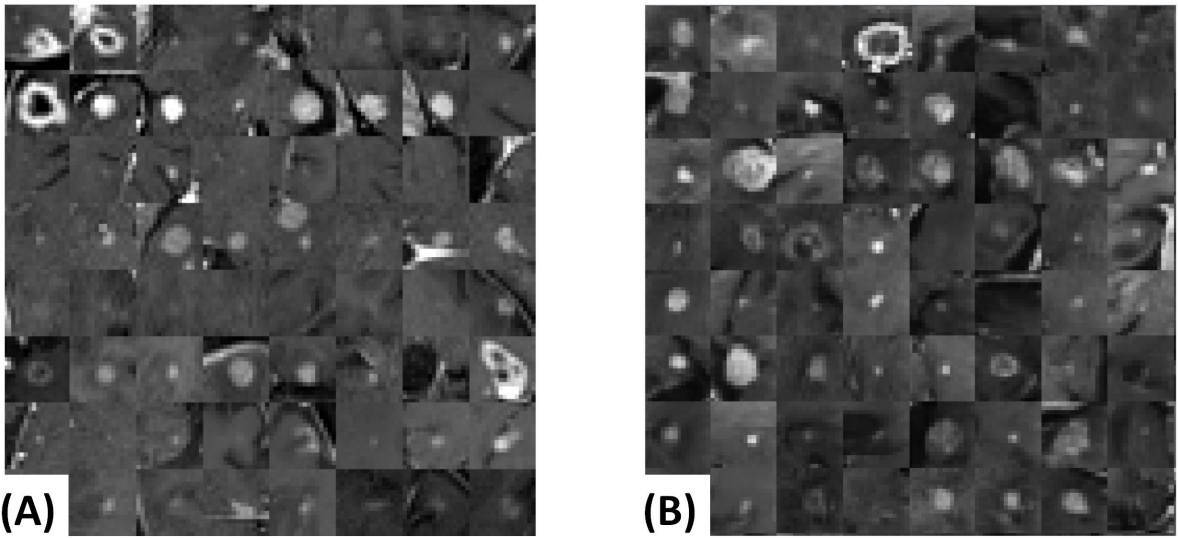

Fig. 3 Mosaic of mid-axial slices of (A) real and (B) DCGAN generated synthetic BM region volumes.

2.2 Ensemble of GANs

The optimization of Equation-1 is proven to minimize the Jensen-Shannon divergence (JSD) between the PDFs of real and synthetic data [18]. As illustrated by Theis et al. [35], the minimization of JSD in a GAN implementation is prone to introducing synthetic data with a PDF that may partially cover the real data's PDF; the modes of the real data may not be fully represented, or even collapsed (i.e., "mode collapse"). Various approaches have been introduced to address/reduce this problem [31,36] including the ensemble of GANs [37]. The method proposes to produce multiple GANs for a given training data, then randomly pick and utilize a GAN generator from this ensemble when a synthetic data is needed. It was validated for collections of DCGANs, and shown to increase the mode coverage as the ensemble size grows.

The reasons for an ensemble approach to improve the decision performance include: (1) avoidance of overfitting due to multiple hypotheses covered by its components, (2) reduced chance of stagnating at a local optima as each component runs its optimization process individually, and (3) improved representation of the optimal hypothesis since the combination of different models commonly expands the solution search space [38].

The performance of an ensemble mainly depends on two factors [39]:

(1) Component performance: Each component (i.e., base-learner) must perform better than a random model; an ensemble of diverged or mode-collapsed GANs would likely fail to produce plausible synthetic samples. Accordingly, we propose to constrain GANs' inclusion into an ensemble with a criterion described in the following section.
(2) Component diversity: The generative diversity of the components is the main reason for an ensemble to perform better than its components [39]. In this study, the framework achieves acceptable levels of diversity by growing into a size controlled by a validation model: For a properly grown ensemble, the validation model that is trained with the synthetic data does not underperform its original-data trained version.

2.3 GAN Inclusion Criterion

The inclusion criterion/criteria are defined for deciding if a given GAN can be included in a cGANe. Various quantitative measures for the evaluation of GANs have been previously described [40–44]. In the introduced framework, the utilized GAN evaluation metric needs to: (1) be implementable for 3D samples, (2) be defined for a single label output (i.e., the cropped BM region), and (3) allow the generation of a cGANe that eventually satisfies the validation criterion/criteria. Based on these, we propose to use the FD [45] between the synthetic and real cropped BM region sample sets;

$$d^2\left((m_r, C_r), (m_g, C_g)\right) = \|m_r - m_g\|_2^2 + Tr\left(C_r + C_g - 2(C_r C_g)^{1/2}\right), \tag{2}$$

where $(m_r, C_r)$ and $(m_g, C_g)$ give real and generated data mean vectors and covariance matrices, respectively. Unlike the Fréchet Inception Distance (FID) [40], where the data's reduced-dimension representation is retrieved from an embedding layer of the Inception network [46], flattened-original (without any downsampling) versions of both the real and synthetic data are represented with multivariate Gaussian distributions in our implementation. Relatively lower scale of our data samples enables this direct usage; $m_r$ and $m_g$ are each defined in $\mathbb{R}^{4096}$, and $C_r$ and $C_g$ are each defined in $\mathbb{R}^{4096} \times \mathbb{R}^{4096}$. Noteworthy is the fact that the Inception network (or any 2D convolutional neural network (CNN))- related GAN measures [40,42,44] cannot be applied in our application; as the generated data is in 3D and has a single problem-specific label.

In the proposed framework, a GAN is versioned throughout its training. In every $n$ epochs of the training,

1. The GAN's current state (i.e. its temporary version) is stored.
2. $M$ synthetic data samples are produced via the generator network of the current GAN version, where each synthetic sample is conditioned to differ from all real training samples as:

$$I_{max} = argmax_{n \in \{1,N\}}\left(H(S_g) - H(S_g|S_{r,n})\right), and\ I_{max} \leq \varphi, \tag{3}$$

with: (1) $N$ is the number of real training samples, (2) $S_g$ is the synthetic sample, (3) $S_{r,n}$ is a real sample, (4) $H(S_g)$ is the Shannon entropy of the synthetic sample, and (5) $H(S_g|S_{r,n})$ is the conditional entropy. Accordingly, $I_{max}$ gives the maximum mutual information (MI) between the synthetic sample and all real samples, and $\varphi$ is the maximum acceptable MI.

3. The FD between the synthetic and real data samples is computed and stored.

After the training is complete, the GAN version producing the lowest FD, which is also lower than a predefined threshold value ($\omega$), is added to the cGANe. Otherwise, the produced GAN is not acceptable. Accordingly, the inclusion criterion for a given GAN is to produce synthetic data samples that differ from the original data; yet, have FD with the original data below a predefined threshold $\omega$. Figure 4 illustrates the computed FD throughout the training of an example DCGAN.

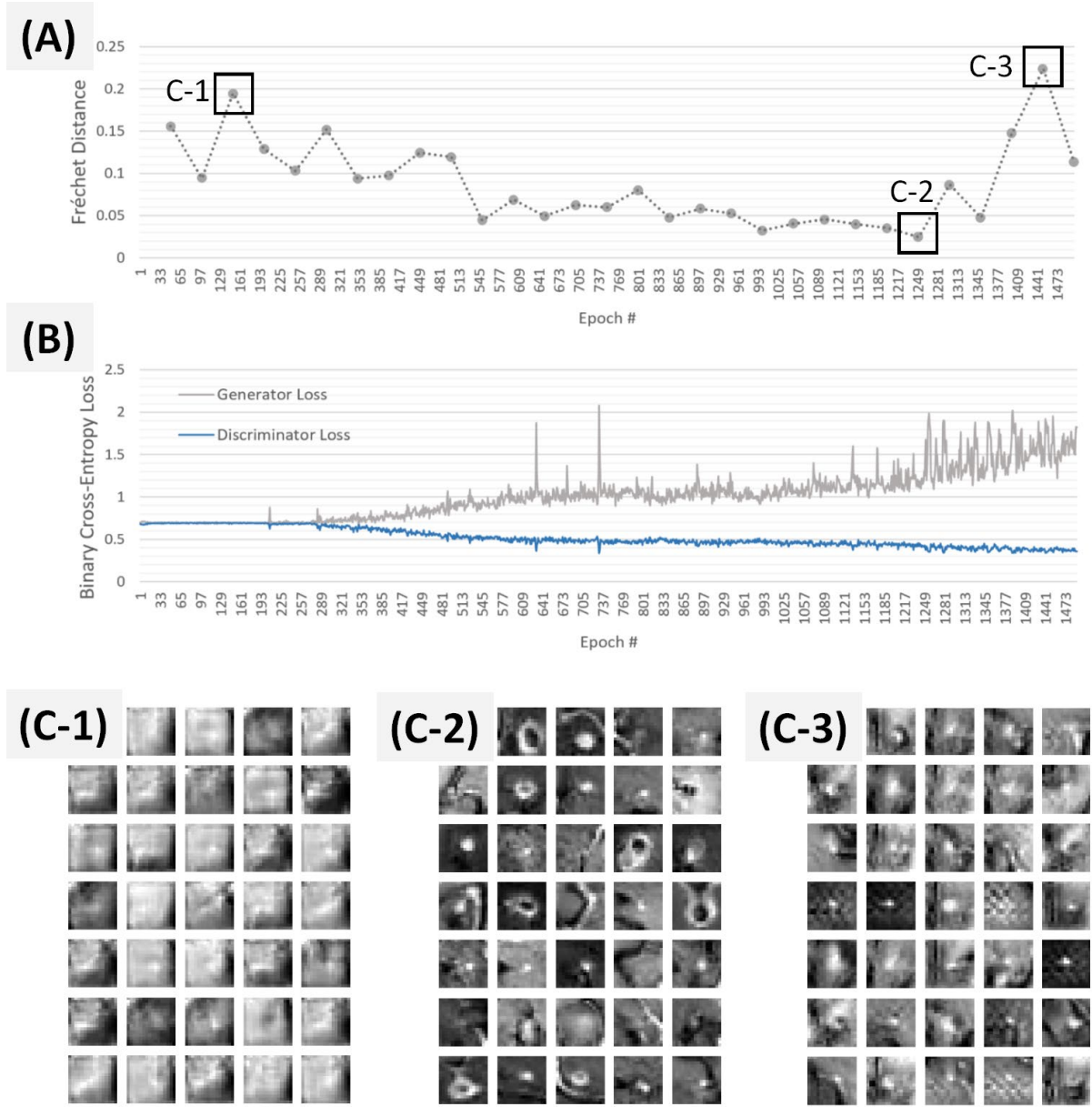

Fig. 4 Fréchet Distance (FD) for the GAN validation.

2.4 Validation Model - Brain Metastases Detection Framework

The validation model is utilized to (1) define baseline performance metrics using the original data, (2) assess the performance of the system when the synthetic data is used during its training, and (3) determine if the generated cGANe is complete or needs a further growth. The strategy adopts the presumption that the validity of synthetic data can be determined by its impact on a system that is pre-evaluated with the original (i.e., non-synthetic) data. Multiple previous studies utilizing GANs for data augmentation compared their results against the version of their system that did not use GANs for data augmentation [22–24]. Visual Turing Test [47], deploying medical experts to grade synthetic images based on their visual appearances, has also been utilized in the given context [24,33]. However, the visual grading of synthetic data may be problematic as small artifacts or abnormalities in a synthetic image (e.g., fixed range of intensities at a specific pixel, fixed intensity ratio between parts of the image, etc.), which might be unnoticeable or downplayed by a human grader, can easily be exploited by a ML model.

The BM-detection framework [27] is used as the validation model for the given case study. Briefly, the framework consists of two main components, consisting of candidate-selection and classification. During its deployment, the input MRI volume is processed using an information theory-based approach for the detection of image points with high probabilities of representing BMs. Next, volumetric regions centered by these candidate locations are iteratively fed into a classification CNN. The model is trained with random paired data batches; each batch includes an equal number of positive and negative volumetric regions (i.e., the positive volumetric region contains a BM, whereas the negative does not). The positive data samples are augmented on the fly [48] using random: (1) Simard type [49] elastic deformations, (2) gamma corrections, (3) image flips, and (3) image rotations (see Figure 5). (Interested readers are referred to [27] for further details on CNN specifications and the augmentation procedures of the detection framework.)

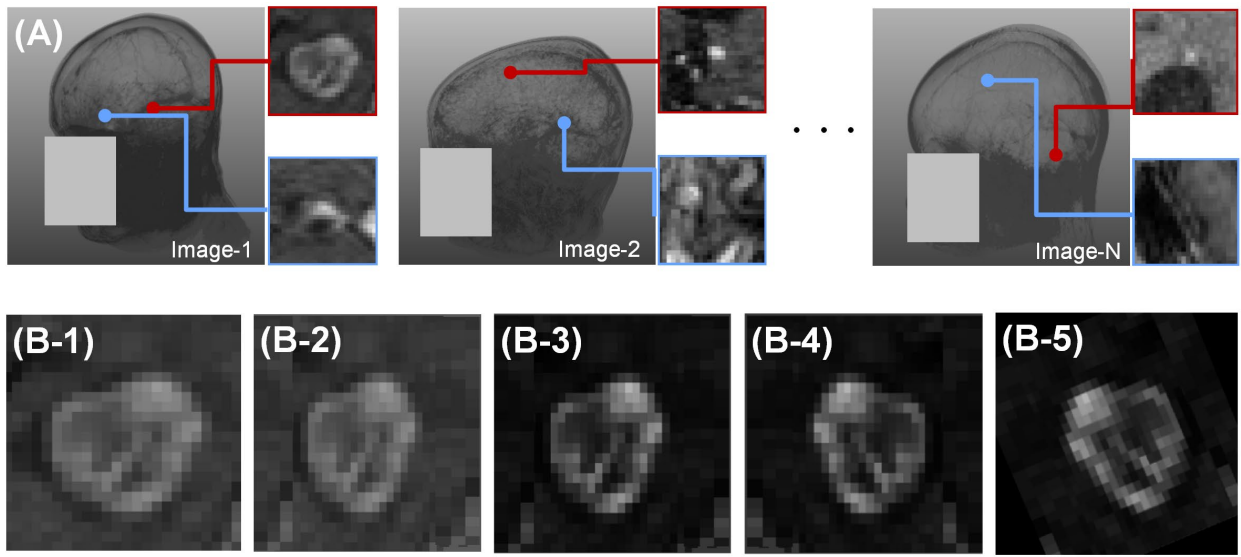

Fig. 5 Training of the BM-detection framework.

# 3. Database

3.1 Data Collection

This retrospective study was conducted under Institutional Review Board approval with a waiver of informed consent. A total of 217 post-gadolinium T1-weighted 3D MRI exams were collected from 158 patients: 113 patients with a single dataset, 33 patients with 2 datasets, 10 patients with 3 datasets, and 2 patients with 4 datasets. The images were collected from 8 scanners, where the acquisition parameters for each are summarized in Appendix-A.

3.2 Brain Metastases

The database included 932 BMs where,

- The mean number of BMs per patient is 4.29 ($\sigma$ = 5.52),
- The median number of BMs per patient is 2,
- The mean BM diameter is 5.45 mm ($\sigma$ = 2.67 mm),
- The median BM diameter is 4.57 mm,
- The mean BM volume is 159.58 mm3 ($\sigma$ = 275.53 mm3),
- The median BM volume is 50.40 mm3.

Histograms for the BM count, BM diameter, and BM volume are provided in Appendix-B.

# 4. Validation Settings

4.1 Validation Metric

The BM-detection algorithm is evaluated by measuring the average number of false detections per patient (AFP) for a range of detection sensitivities, where the metric was adopted by various state-art-of art approaches [50,51]. In the current study, AFP is utilized for:

- The baseline performance calculation: Random paired data batches collected from the training data are fed into the validation model with on the fly augmentations, as described in the previous section (see Figure 6-A). Then, the trained validation model is executed with the test data for producing a baseline-AFP.
- The validation criterion for a cGANe: First, the training data is used for generating the cGANe. During the validation model training, the positive samples of the paired batches are: (1) generated by the cGANe on the fly, and (2) fed into the validation model without any augmentation (see Figure 6-B). Finally, the trained validation model is executed with the test data. If the computed AFP is higher than the baseline-AFP, then the cGANe needs further growth. The convergence of the AFP values to the baseline-AFP shows the adequacy of synthetic samples for the reproduction of the research results elsewhere.

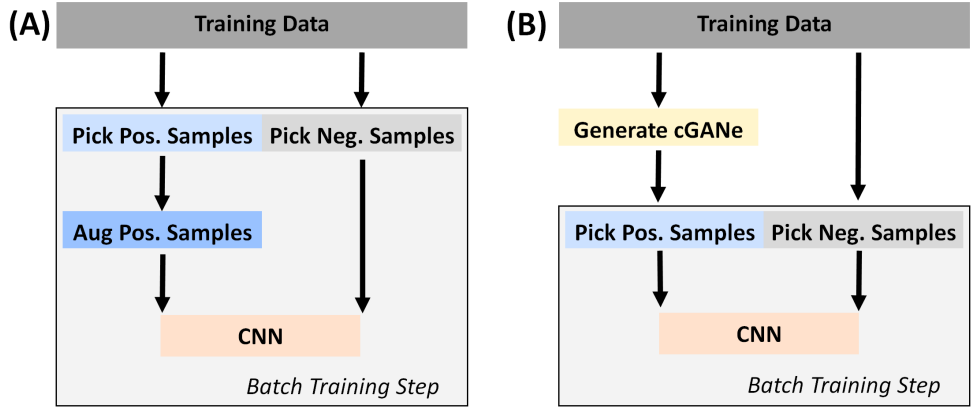

Fig. 6 Training data preparation during the (A) baseline and (B) cGANe setups.

4.2 Data Preprocessing

All datasets were resampled to have isotropic (1mm × 1mm × 1mm) voxels. The voxel values were normalized to [0, 1] range, where the maximum and minimum intensity voxels for each dataset had the normalized values of 1 and 0 respectively.

4.3 System Parameters

Binary-cross entropy was used as the loss function for the discriminator and the combined DCGAN networks (i.e., the generator and the discriminator). The optimizations of the networks were performed via Adam algorithm [52], where the learning rate for the discriminator was set to 0.00005, and it was set to 0.0003 for the DCGAN. The optimizers' exponential decay rates for the first and second-moment estimates were 0.5 and 0.999 respectively. The dropout rate of the discriminator network was 0.15, and leaky ReLU units' alpha values were 0.1 for both of the networks. The DCGAN training was performed in 1500 epochs with the batch sizes of 8 pairs of positive and negative samples.

The FD between the real data and synthetic data was computed in every 50 epochs; (1) the real data statistics (i.e. $m_r$ and $C_r$) were computed using all training data, and (2) the synthetic data statistics (i.e. $m_g$ and $C_g$) were computed using $M = 2000$ random synthetic data samples created by the current state of the generator, where the maximum allowed mutual information between a synthetic sample and real samples was $\varphi = 0.5$. The FD threshold was set as $\omega = 0.04$.

The validation model parameters were adopted from our prior work [27] without any modifications. The cGANe validations were performed for the increments of 10 constrained DCGAN components (e.g., cGANe10, cGANe20, etc.). The validation criterion was set as achieving baseline-AFP ± 1.0 at 90 percent BM-detection sensitivity.

## 5. Results

Both the (1) baseline-AFP and (2) AFP after the usage of cGANe generated synthetic data were computed using 5-fold cross-validation (CV). The data in these folds were patient-wise divided; hence, the folds included 31, 31, 32, 32 and 32 patients' data respectively. For each CV fold, four folds were used for the training and one fold was used for the testing, where the data in the test fold was kept unmodified (i.e., no augmentations were applied, and no synthetic data was derived from it). For the baseline, cGANe10, cGANe20, cGANe30 and cGANe40 setups, the AFP values in connection to the detection sensitivity are displayed for each CV fold with their mean in Figure 7. For these setups, the AFP values for 75, 80, 85 and 90 sensitivity percentages are reported in Table 1.

TABLE 1
AFP VS SENSITIVITY

| Sensitivity % | Baseline | cGANe10 | cGANe20 | cGANe30 | cGANe40 |
|---|---|---|---|---|---|
| 75 | 1.90 | 10.60 | 6.62 | 4.26 | 3.19 |
| 80 | 2.96 | 12.82 | 9.56 | 6.24 | 4.32 |
| 85 | 5.85 | --- | 13.42 | 8.26 | 6.22 |
| 90 | 9.12 | --- | --- | 12.47 | 9.53 |

Average number of false positives at specific sensitivity percentages are reported for the baseline, cGANe10, cGANe20, cGANe30 and cGANe40 setups.

The proposed solution was implemented using the Python programming language (v3.6.8). The neural network implementations were performed using Keras library (v2.1.6-tf) with TensorFlow (v1.12.0) backend. The training of each DCGAN was done in ~1.25 hours, where a DCGAN satisfying the inclusion criterion was generated in ~2.15 hours on average. Thus, growing a given cGANe with ten additional DCGANs took ~21.5 hours on average. The training of the validation model for each fold took ~3.5 hours [27]. The network training was performed using four parallel processing NVIDIA 1080ti graphics cards, having 11 GB RAM each.

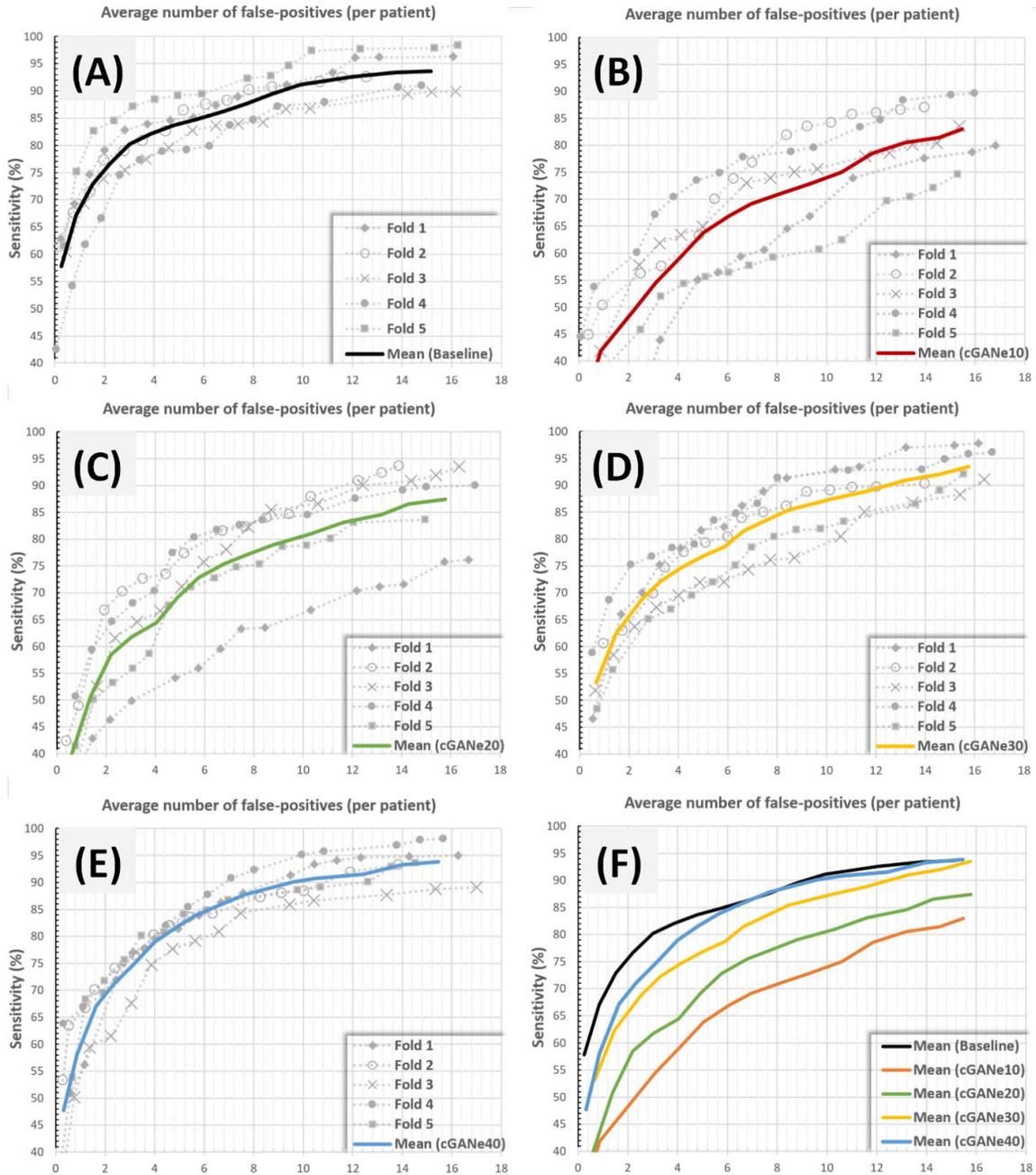

Fig.7. AFP in relation to the detection sensitivity for the (A) baseline, (B) cGANe10, (C) cGANe20, (D) cGANe30 and (E) cGANe40. (F) The average curves for the baseline and cGANe setups.

## 6. Discussion

The results show that the cGANe AFP value reduces with the number of constrained DCGANs it includes; cGANe30 and cGANe40 produced AFP values of 12.47 and 9.53 for 90 percent sensitivity respectively. As the validation criterion for cGANe was set as producing baseline-AFP ± 1.0 at this detection sensitivity, which is 9.12 ± 1.0, cGANe40 passed this stage by producing AFP of 9.53. The results suggest that the usage of cGANe40-generated positive synthetic samples can lead to a model that produces comparable results with the original validation model (i.e., trained with real positive samples and their augmentations); indicating a highly significant finding. Accordingly, the ensemble can be utilized for producing positive synthetic data samples for client sites intending to (1) reproduce the results with the same BM-detection model, or (2) use it for performing a closely related research requiring this specific data type (i.e., 3D brain metastases on T1-weighted contrast-enhanced MRI examinations).

As described previously, a potential problem with the usage of a single GAN is the partial representation of the real data PDF. The issue and the validity of our solution was further illustrated by performing a low dimensional data embedding analysis (see Figure 8): The real data (i.e., all 932 BMs) and the matching number of cGANe generated synthetic samples were visualized via two-dimensional embeddings, generated by (1) reducing the flattened 4096-dimensional volumetric data into 80-dimensional data using principal component analysis (PCA) [53], explaining ~84.5 percent of the data variance, and (2) embedding these 80-dimensional representations into two dimensions using t-Distributed Stochastic Neighbor Embedding (t-SNE) [54]. (The mapping of very high dimensional data into highly representative lower-dimensional data prior to t-SNE was suggested in [54]). As shown in the cGANe1 plot, the usage of a single constrained DCGAN caused the lower-dimensional mappings to accumulate in regions that do not align well with the original data. The misrepresentation declined with the cGANe scale, where the cGAN(e≥10) plots have better real and synthetic data mixtures; explaining the improved validation model performances of cGANe settings with higher numbers of components.

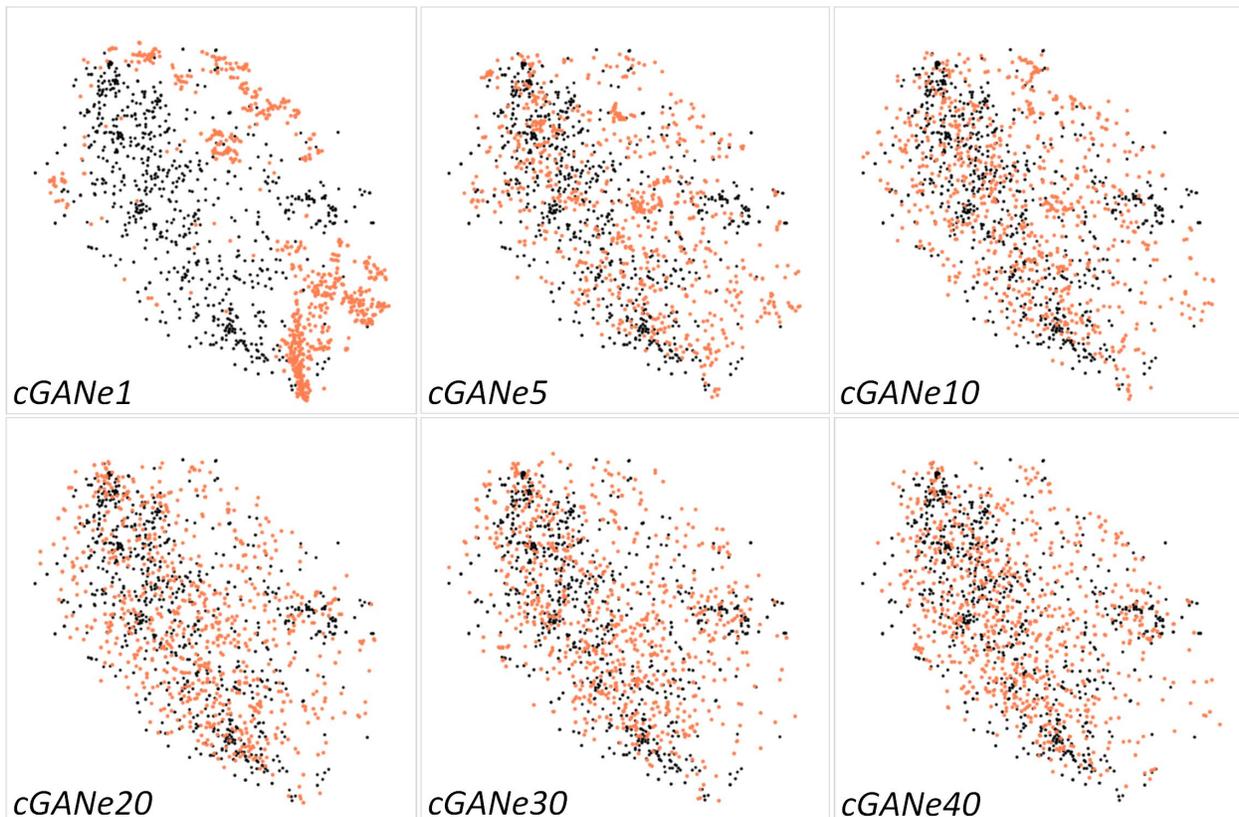

Fig. 8 t-SNE representations for real (black) and cGANe generated (orange) data samples.

The introduced framework may be modified for a variety of applications in a multitude of fields (e.g., medicine, finance, e-commerce, etc.) with limitations regarding the sharing of sensitive data. The framework should be adapted by first deciding on a validation model that the clients are interested in using. Next, a GAN formulation needs to be determined that: (1) fits the data type (e.g., PGGAN [32] for high-resolution 2D images, RC-GAN [55] for time series, etc.), and (2) produces visually pleasing results. Accordingly, a proper GAN inclusion criterion needs to be set to verify the outputs of the chosen GAN; for instance, the FID may be a proper choice for 2D image-producing GAN formulations. After these major components are defined, cGANe can be grown to a proper size, in which the synthetic data allows the validation model to achieve validation criteria that were set using the original data. The wide-range of adaptations of the introduced framework for different applications may be investigated in future studies.

However, the cGANe may fail to pass the final validation stage after several growth steps. This might be due to many reasons, including (1) improper framework components, such as inappropriate GAN formulation for a given data type or suboptimally set GAN inclusion constraint, and (2) unachievable validation criteria. The first item may be addressed by experimenting with various GANs and relevant group of constraints. The validation criteria, which are derived from the validation model using the original data and its augmentations, need to consider the fact that GAN alternates the representation of the information with some information gap-filling; hence, it does not introduce novel information that significantly improves the system performance [23]. Accordingly, the validation criteria may not be over-ambitious, such as significantly outperforming non-synthetic data using version of the validation model.

The mutual information (MI) based metric, defined in Equation-3, aims to eliminate the generation of a synthetic data sample that resembles an original data sample from the training set. As the metric utilizes a joint intensity histogram between the synthetic and original data samples during its computation [56], the translation or rotation of one of these samples may cause the MI to decrease, even if the samples are identical. In this study, the problem was dealt with by keeping the maximum acceptable MI relatively low ($\varphi = 0.5$), eliminating small portion of the novel synthetic data samples, whereas increasing the time during the GAN inclusion criterion. In a future study, an additional data-alignment procedure [57], may be used prior to the MI computation to improve this aspect of the framework.

The study introduced a synthetic data-generation framework with the purpose of sharing the data with other sites. The technical components of the framework are well-known algorithms that are adapted and combined for a sound and novel workflow; they include the following: (1) DCGAN approach is generalized to 3D, (2) FD is used for the whole data representation taking advantage of the low-scale data, without down-sampling to feature domain such as FID, (3) the mutual information-based metric is defined to ensure that the synthetic data is not identical to the original data, and (4) a previously reported BM-detection framework [27] is utilized as the validation model to test the viability and information coverage of the synthetic data. The results suggest that the cGANe generated synthetic data holds great potential to replace the original data as long as the ensemble components are constrained and the ensemble size is controlled using a validation model.

## Appendix A – Scanner Parameters

TABLE 2
SCANNER PARAMETERS

| Scanner | MF [a] (T) | TR [b] range (ms) | TE [c] range (ms) | Slice thickness range (mm) | Pixel size [d] range (mm) | Imaging frequency range (MHz) | Flip angle range (degrees) | Exam # |
|---|---|---|---|---|---|---|---|---|
| Siemens Aera [e] | 1.5 | [9.3, 9,7] | [4.4, 4.7] | 1.0 | [0.78, 0.97] | 63.6 | 20 | 54 |
| Siemens Avanto [e] | 1.5 | [9.7, 10] | [4.2, 4.8] | [0.9, 1.0] | [0.43, 0.86] | 63.6 | [15, 20] | 17 |
| Siemens Espree [e] | 1.5 | 10 | [4.5, 4.7] | 1.0 | [0.78, 1.0] | 63.6 | 20 | 26 |
| Siemens Skyra [e] | 3.0 | [6.2, 6.5] | 2.46 | [0.8, 0.9] | [0.65, 0.78] | 123.2 | [10.5, 12] | 34 |
| Siemens TrioTrim [e] | 3.0 | 6.5 | 2.45 | [0.8, 0.9] | [0.65, 0.73] | 123.2 | 10.5 | 4 |
| Siemens Verio [e] | 3.0 | [6.5, 9.0] | [2.4, 4.9] | [0.8, 0.9] | [0.65, 0.78] | 123.2 | 10.5 | 28 |
| GE Optima MR450w [f] | 1.5 | 10.3 | 4.2 | 1.0 | 0.49 | 63.9 | 20 | 26 |
| GE Signa HDxt [f] | 1.5 | [9.2, 10.3] | 4.2 | 1.0 | [0.49, 0.98] | 63.9 | 20 | 28 |

[a] Magnetic field strength, [b] repetition time, [c] echo time, [d] pixel size is same in x and y directions.
[e] Siemens Healthcare, Erlangen, Germany.
[f] GE Healthcare, Milwaukee, Wisconsin, USA.

## Appendix B – BM Histograms

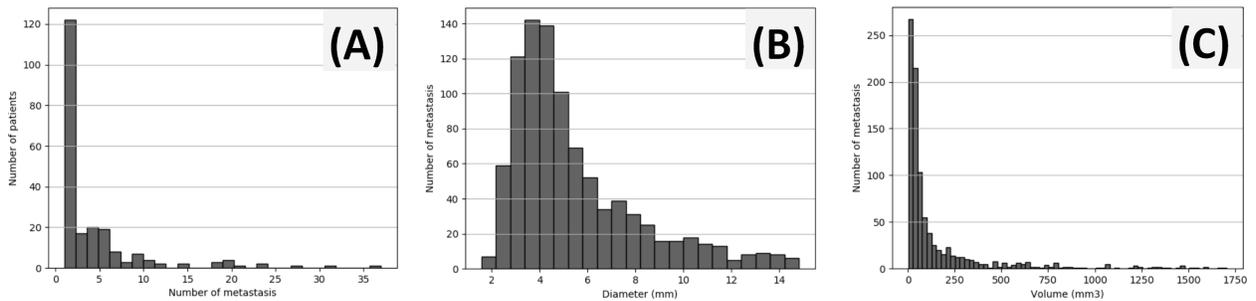

Fig. 8 Histograms for the BM (A) count per patient, (B) diameter and (C) volume.

## Caption List

**Fig. 1** Constrained GAN ensemble for data sharing: (1) Modality generates the original images, (2-A) the original data is used for training a GAN, (2-B) validation model computes baseline accuracy, (3) the GAN is tested with inclusion criteria, (4) if the inclusion criteria are met, then GAN is included in a GAN ensemble, (5) synthetic data is generated with the ensemble, (6) the synthetic data is validated with the validation model, (7) if the data passes validation criteria, then the ensemble is complete, otherwise ensemble needs to grow further, (8) validated GAN ensemble is used for generating synthetic data for other sites.

**Fig. 2** The generator (A) and discriminator (B) networks of the used 3D DCGAN: Classical contracting and expanding architectures are deployed with 3D convolution layers.

**Fig. 3** Mosaic of mid-axial slices of (A) real and (B) DCGAN-generated synthetic BM region volumes.

**Fig. 4** Fréchet Distance (FD) for the GAN validation: (A) FD between the original and sample set of generated data were computed periodically, in every 50 epochs; the minimal distance was reported at C-2 measurement point. (B) Binary cross-entropy loss of the GAN's generator and discriminator are provided as a reference. (C) For three reference points (i.e. C-1, C-2 and C-3), mid-axial slices of randomly generated BM region volumes are shown: In C-1 and C-3, the samples do not resemble real BM appearance; C-1 presents limited variability, and C-3 has multiple samples with checkerboard like artifacts. In C-2, the samples resemble the actual BM appearances; they are in various dimensions/contrasts, some even have cystic formations.

**Fig. 5** Training of the BM-detection framework: (A) Positive (red) and negative (blue) pairs of volumetric BM regions are randomly collected for each batch. (B) Each positive BM region volume goes through random augmentations; mid-axial slice (B-1) of the original cropped volume, (B-2) after random elastic deformation, (B-3) after random gamma correction, (B-4) after random flipping, and (B-5) after random rotation.

**Fig. 6** Training data preparation during the (A) baseline and (B) cGANe validation setups.

**Fig. 7** AFP in relation to the detection sensitivity for the (A) baseline, (B) cGANe10, (C) cGANe20, (D) cGANe30, (E) cGANe40. (F) The average curves for the baseline and cGANe setups.

**Fig. 8** t-SNE representations for real (black) and cGANe generated (orange) data samples.

**Fig. 9** Histograms for the BM (A) count per patient, (B) diameter and (C) volume.

## Funding

This research did not receive any specific grant from funding agencies in the public, commercial, or not-for-profit sectors.